\documentclass[ALICE,manyauthors]{cernphprep}
\usepackage{aliprep}
\usepackage[numbers,sort&compress]{natbib}
\usepackage{textcomp}
\usepackage[T1]{fontenc}
\usepackage{lineno}

\newcommand{\sqrtsNN}{\sqrt{s_{\scriptscriptstyle \rm NN}}}
\newcommand{\lsim}{\,{\buildrel < \over {_\sim}}\,}
\newcommand{\gsim}{\,{\buildrel > \over {_\sim}}\,}
\newcommand{\av}[1]{\left\langle #1 \right\rangle}

\newcommand{\gev}{\mathrm{GeV}}
\newcommand{\gevc}{\mathrm{GeV}/c}
\newcommand{\tev}{\mathrm{TeV}}

\newcommand{\Raa}{R_{\rm AA}}
\newcommand{\fnonprompt}       {f_\mathrm{non\text{-}prompt}}
\newcommand{\rawY}[1]          {Y_{#1}}
\newcommand{\effNP}[1]         {(\mathrm{Acc}\times\epsilon)^\mathrm{non\text{-}prompt}_{#1}}
\newcommand{\effP}[1]          {(\mathrm{Acc}\times\epsilon)^\mathrm{prompt}_{#1}}
\newcommand{\Nnp}              {N_\mathrm{non\text{-}prompt}}

\begin{document}

\begin{titlepage}
\PHyear{2022} 
\PHnumber{015}
\PHdate{28 January}

\title{Measurement of beauty production via non-prompt $\Dzero$ mesons in $\PbPb$ collisions at \fNN}
\ShortTitle{Beauty production via non-prompt $\Dzero$ in $\PbPb$ collisions at \fNN}

\Collaboration{ALICE Collaboration\thanks{See Appendix~\ref{app:collab} for the list of collaboration members}}
\ShortAuthor{ALICE Collaboration}


\begin{abstract}
The production of non-prompt $\rm D^0$ mesons from beauty-hadron decays was measured at midrapidity ($\abs{y} < 0.5$) in Pb--Pb collisions at a nucleon--nucleon center-of-mass energy of $\sqrt{s_{\text{NN}}}=5.02~\tev$ with the ALICE experiment at the LHC. Their nuclear modification factor ($\RAA$), measured for the first time down to $p_{\rm T}=1~\gev/c$ in the \cent{0}{10} and \cent{30}{50} centrality classes, indicates a significant suppression, up to a factor of about three, for $\pT > 5~\gev/c$ in the 0--10\% central Pb--Pb collisions. 
The data are described by models that include both collisional and radiative processes in the calculation of beauty-quark energy loss in the quark--gluon plasma, and quark recombination in addition to fragmentation as a hadronisation mechanism. 
The ratio of the \np{} to prompt $\Dzero$-meson $\RAA$ is larger than unity for $\pT > 4~\gev/c$ in the 0--10\% central Pb--Pb collisions, as predicted by models in which beauty quarks lose less energy than charm quarks in the quark--gluon plasma because of their larger mass.

\end{abstract}

\end{titlepage}

\setcounter{page}{2}
\section{Introduction}
\label{sec:intro}
The formation in ultra-relativistic collisions of heavy nuclei of a quark--gluon plasma (QGP), a state in which quarks and gluons are not confined into hadrons, is supported by several measurements at SPS, RHIC and LHC accelerators~\cite{NA50:2000brc, WA97:1999uwz,BRAHMS:2004adc,PHENIX:2004vcz,PHOBOS:2004zne,STAR:2005gfr,Schutz:2011zz,Roland:2014jsa,Braun-Munzinger:2015hba}, and expected from quantum chromodynamics (QCD) on the lattice~\cite{Karsch:2006xs,Borsanyi:2010bp,Borsanyi:2013bia,Bazavov:2011nk}. Heavy quarks (charm and beauty) 
are produced in hard-scattering processes 
occurring in the early stage of the collision. 
As the medium expands, they interact with its constituents via inelastic (gluon radiation)~\cite{Gyulassy:1990ye,Baier:1996sk} and elastic~\cite{Thoma:1990fm,Braaten:1991jj,Braaten:1991we} flavour-conserving scatterings that modify their momentum towards equilibrium with the surrounding quarks and gluons. As a consequence, high-momentum charm and beauty quarks lose energy. This in-medium energy loss, which carries information on the medium properties and expansion, can be investigated by measuring the nuclear modification factor ($\Raa$) of heavy-flavour hadrons. The $\Raa$ is defined as the ratio of the transverse-momentum ($\pt$)-differential production yields measured in a given centrality interval in nucleus--nucleus collisions (${\rm d} N_{\rm AA}/{\rm d}\pt$) to the cross section in proton--proton (pp) collisions (${\rm d}\sigma_{\rm pp}/{\rm d}\pt$) scaled by the average nuclear overlap function $\av{T_{\rm AA}}$~\cite{Glauber:1970jm,ALICE:2018tvk} in the considered centrality interval. In-medium energy loss of quarks and gluons implies a softening of the final-state hadron $\pt$ spectrum resulting in $\Raa<1$ at intermediate and high $p_{\rm T}$.
Several measurements in Au--Au collisions at RHIC and Pb--Pb collisions at the LHC evidence a substantial energy loss of charm and beauty quarks due to their interactions in the QGP~\cite{Acharya:2018upq,Acharya:2018hre,ALICE:2012ab,Abelev:2012qh,Sirunyan:2017xss,Adam:2015jda,Adam:2015sza,Adam:2016wyz,ALICE:2021rxa,ALICE:2021kfc,Adamczyk:2014uip,Abelev:2006db,Adare:2012px,Adare:2010de,Adler:2005xv,CMS:2015sfx,CMS:2012bms,CMS:2016mah,CMS:PhysRevLett.123.022001,CMS:2017uoy,Aaboud:2018bdg,Aaboud:2018quy}. The difference between the $\RAA$ of heavy-flavour hadrons or their decay products and that of light hadrons, mostly originating from gluon and light-quark fragmentation, indicates that the amount of energy loss is sensitive to the colour-charge dependence of the strong interaction, as well as to the effects that depend on the parton mass~\cite{ALICE:2015dtd,ALICE:2021rxa,CMS:2017uuv,Adam:2015nna}. 
In particular, beauty quarks are expected to lose less energy than charm quarks.
At high $\pt$, where energy loss is caused mainly by radiative processes, 
this difference is expected to derive from the ``dead-cone" effect, which suppresses gluon radiation off massive quarks at angles smaller than $m_{\rm Q}/E$ (with $m_{\rm Q}$ and $E$ being the quark mass and energy, respectively) with respect to the quark direction~\cite{Dokshitzer:2001zm,Armesto:2003jh,Zhang:2003wk,Djordjevic:2003zk}, an effect directly observed in pp collisions at the LHC~\cite{ALICE:2021aqk}. This expectation is supported by experimental data showing higher $\Raa$ for beauty than charm signals, qualitatively in line with theoretical predictions~\cite{STAR:2021uzu,PHENIX:2015ynp,ALICE:2021rxa,CMS:2017uuv,Adam:2015nna,ATLAS:2021xtw,Aaboud:2018quy,CMS:PhysRevLett.123.022001,ALICE:2016uid,ALICE:2015nvt}. At low momenta, heavy-quark propagation through the medium is described as a diffusion process, occurring via multiple low-energy-transfer interactions, that also favours the participation of heavy quarks in the collective expansion of the system~\cite{Batsouli:2002qf,Greco:2003vf}. Because of the larger mass, beauty quarks should diffuse less than charm quarks and have a longer relaxation time, which should increase linearly with the quark mass~\cite{Moore:2004tg,Petreczky:2005nh}. Therefore, the comparison of charm and beauty $\RAA$ provides a handle to constrain the modeling of the diffusion process. 
 
Other effects are also relevant in nuclear collisions, 
namely cold-nuclear-matter (CNM) effects, that are present even without the formation of a QGP, and hadronisation effects. The main CNM effect at LHC energies is the modification of the parton distribution functions (PDF), in particular the reduction of the gluon PDF at small Bjorken-{\it x} values (``nuclear shadowing") that can cause a suppression of heavy-flavour production. At midrapidity, shadowing is expected to be relevant mainly at low $\pt$ (below 2--3 GeV/$c$) and stronger for charm than beauty quarks, as suggested also from measurements performed in p--Pb collisions~\cite{Adam:2016wyz,Acharya:2019mno,Acharya:2018yud,Aaij:2017cqq,Aaij:2017gcy,Aaij:2019lkm,Kusina:2017gkz,Aad:2015ddl,Khachatryan:2015uja}.
In a high quark-density environment like the QGP, low-momentum heavy quarks may hadronise by recombining with other quarks in the medium~\cite{Greco:2003vf,Andronic:2003zv}. Such a ``coalescence" mechanism can enhance the production of heavy-flavour baryons and of hadrons with strange quarks relative to non-strange B and D mesons~\cite{Fries:2003vb,Greco:2003vf,Ravagli:2007xx,ALICE:2021kfc,CMS:2018eso} and it influences the $\pt$ and azimuthal distributions of the produced heavy-flavour hadrons in a different way with respect to ``vacuum-like" fragmentation. Several theoretical models need to include this mechanism to describe the measured $\Raa$ and azimuthal anisotropy of D mesons~\cite{He:2014cla,Nahrgang:2013xaa,Li:2019wri,Beraudo:2014boa,Beraudo:2018tpr,Song:2015ykw,Cao:2016gvr,Cao:2017hhk,Ke:2018tsh,Ke:2018jem,Prado:2016szr,Katz:2019fkc}. Hadronisation and CNM effects complicate the determination of fundamental parameters, such as the spatial diffusion coefficient and charm-quark relaxation time, determined from open-charm measurements~\cite{Rapp:2018qla,Li:2019lex}. 

The different impact of the aforementioned effects on beauty- and charm-hadron observables, ultimately due to the different quark masses, offers a handle to constrain these effects and understand heavy-quark diffusion in the medium. 
Moreover, as stated in Ref.~\cite{Rapp:2018qla}, from a theoretical point of view, beauty hadrons represent a cleaner probe of the QGP compared to charm hadrons, in terms of the implementation of both microscopic interactions and transport, and as a measure of coupling strength for quarks that are unlikely to reach thermalisation in the medium~\cite{Rapp:2021dob,Francis:2015daa,Liu:2016ysz,Riek:2010fk}. While several measurements of charm $\Raa$ and azimuthal anisotropy have been performed down to low $\pt$~\cite{Acharya:2018upq,Acharya:2018hre,Abelev:2012qh,ALICE:2015nvt,Sirunyan:2017xss,CMS:2016mah,Adam:2015sza,Adamczyk:2014uip,Aaboud:2018bdg}, the experimental information is still poor for low-momentum beauty hadrons. Existing data on the production of B mesons~\cite{CMS:2017uoy}, J/$\psi$ from beauty decays~\cite{CMS:2017uuv,ALICE:2015nvt,Aaboud:2018quy}, and single leptons from beauty decays~\cite{ALICE:2016uid,ATLAS:2021xtw} are not sensitive to B mesons with $\pt$ around the B-meson mass or lower (for the lepton case the correlation between the lepton and parent beauty-hadron $p_{\rm T}$ is very broad). This leaves unconstrained a kinetic window fundamental to explore the effects mentioned above.

In this paper, we report the measurement of the $\pt$-differential yield and the $R_{\rm AA}$ of $\Dzero$ mesons from beauty-hadron decays (referred to as non-prompt $\Dzero$ mesons) in Pb--Pb collisions at a center-of-mass energy per nucleon pair $\sqrtsNN$ $=$ 5.02~$\tev{}$, for the first time down to $\pt=1~\gevc$ in central (0--10\%) and semi-central (30--50\%) collisions. This represents a significant extension of the previous measurement by CMS~\cite{CMS:PhysRevLett.123.022001} that allows us to compute for the first time the $\pt$-integrated yield of non-prompt $\Dzero$ mesons. The non-prompt $\Dzero$-meson $R_{\rm AA}$ is compared to that of prompt $\Dzero$ mesons,  which are produced in the hadronisation of charm quarks or from the decay of excited open-charm and charmonium states. In what follows, when mentioning a given hadron species we implicitly refer also to its antiparticle.

The paper is structured as follows. The experimental apparatus and data sample used for the analysis are briefly presented in Section~\ref{sec:Det}. The reconstruction of non-prompt $\Dzero$ mesons and all corrections applied to the raw yield are presented in Section~\ref{sec:ana}. The estimation of systematic uncertainties is briefly discussed in Section~\ref{sec:sys}. The results are presented in Section~\ref{sec:resul} and conclusions are drawn in Section~\ref{sec:con}.


\section{Experimental apparatus and data sample}%
\label{sec:Det}
The data were collected with the ALICE detector during the LHC Run 2 in 2018. A detailed description of the ALICE apparatus and of its performance can be found in Refs.~\cite{Aamodt:2008zz,Abelev:2014ffa}. The Time Projection Chamber (TPC)~\cite{Alme:2010ke} is the main tracking device for the measurement of particle momenta. The Inner Tracking System (ITS)~\cite{Aamodt:2010aa} is exploited for the reconstruction of the primary interaction vertex and of the secondary decay vertices of charm- and beauty-hadron decays.
Particle identification (PID) is provided by the measurement of the specific energy loss ${\rm d}E/{\rm d}x$ in the TPC and of the flight time of charged particles from the interaction point to the Time-Of-Flight detector (TOF)~\cite{Akindinov:2013tea}. These detectors, which cover the pseudorapidity interval $|\eta|<0.9$ and full azimuthal angle, are enclosed in a large solenoidal magnet providing a uniform magnetic field of 0.5~T parallel to the LHC beam direction. 
The event triggers and offline selection criteria are defined in Ref.~\cite{ALICE:2021rxa}.  About $1.0\times10^{8}$ and $8.5\times10^{7}$ events in the 0--10\% and 30--50\% centrality classes were selected for further analysis, corresponding to integrated luminosities ($L_{\text{int}}$) of about $130$~\textmu $\rm b^{-1}$ and $56$~\textmu $\rm b^{-1}$, respectively~\cite{ALICE:2018tvk}.

The correction factors for the detector acceptance and the signal reconstruction and selection efficiency were obtained by means of Monte Carlo (MC) simulations. In order to describe the charged-particle multiplicity and detector occupancy, underlying Pb--Pb events at $\sNN=5.02$~TeV were simulated with the HIJING v1.383 generator~\cite{Wang:1991hta}. In order to enrich the simulation of prompt and non-prompt $\Dzero$-meson signals, pp collisions containing a $\rm c\overline c$ or $\rm b\overline b$ pair in each event were simulated with the  PYTHIA 8.243 event generator~\cite{Sjostrand:2014zea} and the particles originating from a charm or a beauty quark were embedded into the underlying Pb--Pb event. The $\pt$ distribution of prompt D mesons in the MC simulation was weighted in order to match the shape measured in data for \mbox{prompt $\rm D^0$ mesons}, while, for non-prompt $\rm{D^0}$ mesons, the parent beauty-hadron $p_{\rm T}$ shape was weighted to match the shape given by model calculations~\cite{Cacciari:1998it,Cacciari:2001td,He:2014cla}.
The generated particles were transported through the apparatus, which was modelled in the simulation using the GEANT3 transport code~\cite{Brun:1994aa}.


\section{Data analysis}%
\label{sec:ana}
The $\Dzero$ mesons were reconstructed via the decay channel ${\rm D}^0 \to {\rm K}^-\pi^+$ with a branching ratio (BR) equal to $(3.950 \pm 0.031)\%$~\cite{Zyla:2020zbs}. The candidates were defined by combining pairs of tracks with opposite charge, each with $|\eta| < 0.8$, $\pt> 0.5$ $(0.4)$$~\gev/c$ for the 0--10\% (30--50\%) centrality class,  a number of crossed TPC pad rows larger than 70 (out of 159), and a minimum number of two hits (out of six) in the ITS, with at least one in either of the two innermost layers, as the main selections.

To reduce the combinatorial background and separate the prompt and non-prompt contributions, a two-step machine-learning classification based on the Boosted Decision Tree (BDT) algorithm provided by the TMVA library~\cite{Hocker:2007ht} was utilised. Variables sensitive to the typical topology of the prompt and non-prompt decay vertices were chosen as input for the BDT algorithm, similarly to what is described in more detail in Ref.~\cite{Acharya:2021cqv}. Before the training, a $\pm 3\,\sigma$ selection around the expected mean ${\rm d}E/{\rm d} x$ in the TPC and time of flight in the TOF was applied to identify pions and kaons, where $\sigma$ is the resolution on the measured quantities. The BDT algorithm was trained in each $\pT$ interval, 
using samples of non-prompt and prompt ${\rm D}^0$ mesons from the MC simulation, and a sample of background candidates with an invariant mass in the sidebands of the $\Dzero$-meson peak from the data. 
In the first step the BDT was trained to separate non-prompt and prompt ${\rm D}^0$ mesons, while in the second step it was trained to separate non-prompt ${\rm D}^0$ mesons and combinatorial background. By tuning the selection on the BDT outputs, the fraction of non-prompt $\Dzero$ can be varied from about 5\% up to 90\% maintaining a reliable signal extraction. 

\begin{figure}[h!]
\begin{center}
\includegraphics[width=0.45\textwidth]{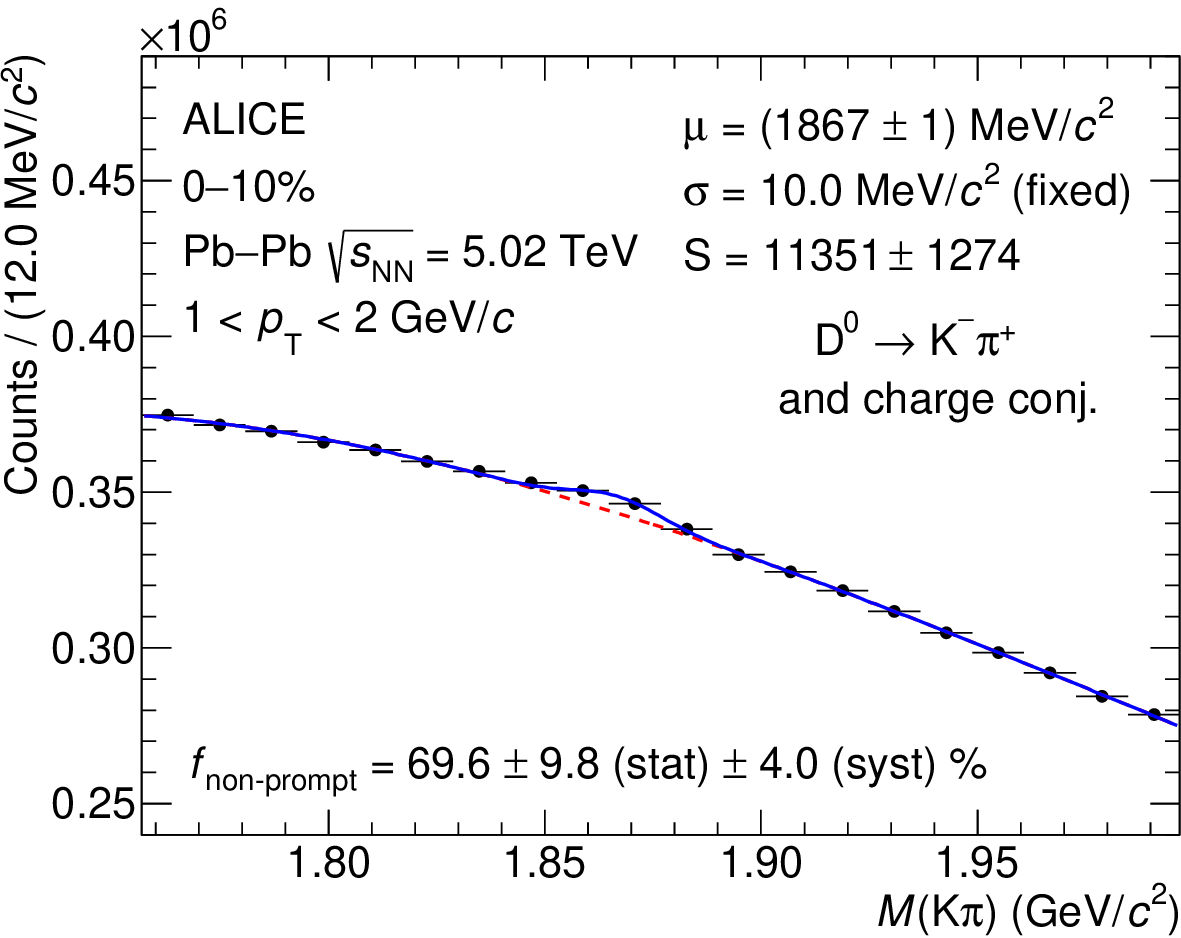}
\includegraphics[width=0.45\textwidth]{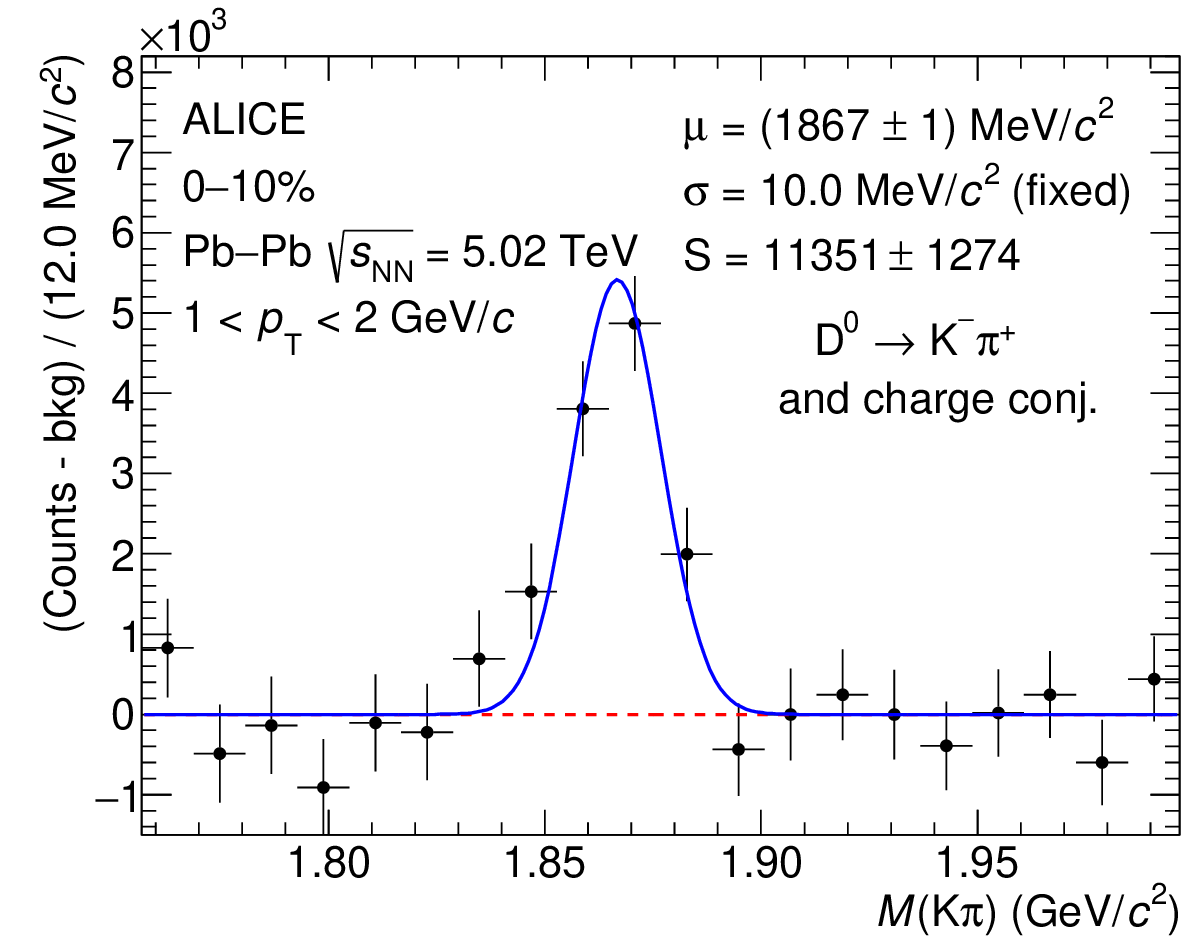}
\includegraphics[width=0.45\textwidth]{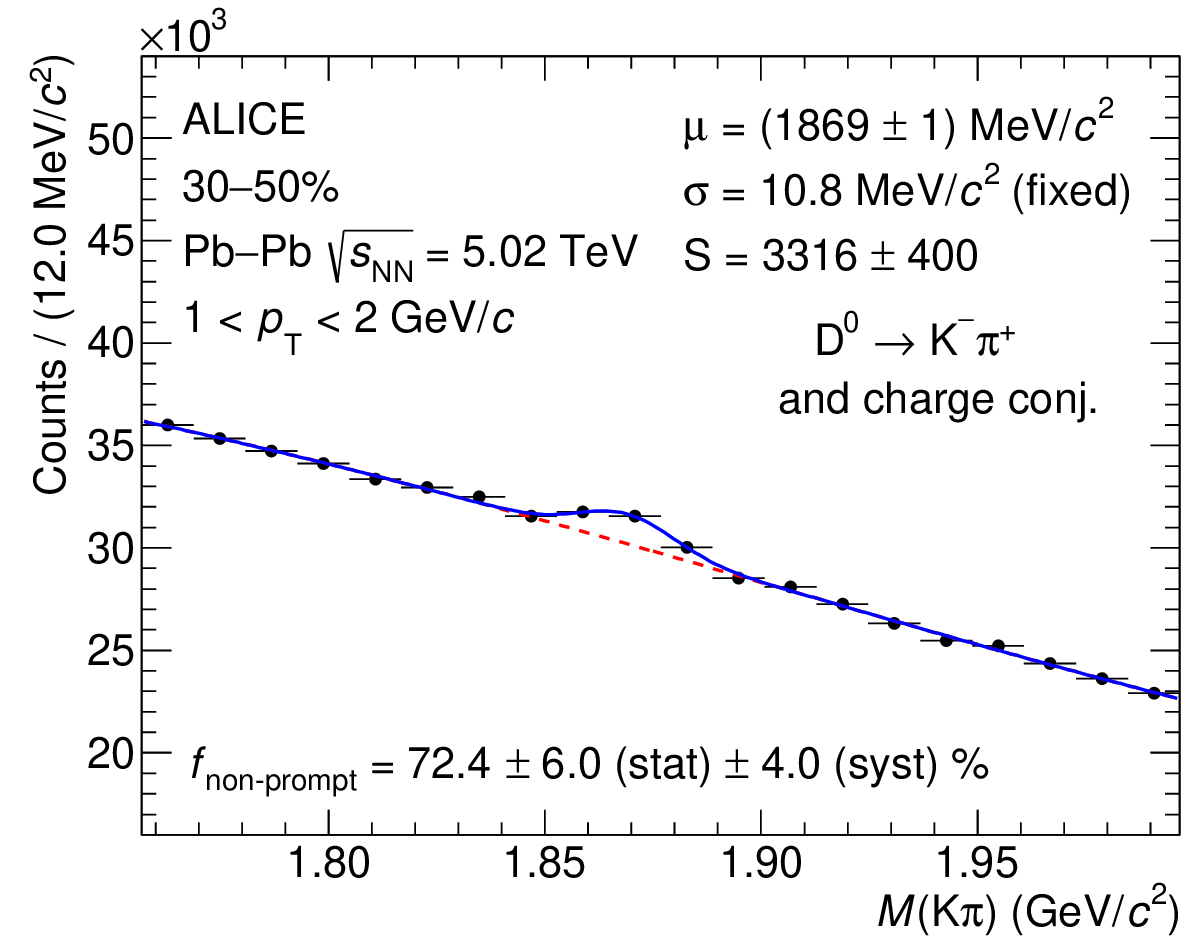}
\includegraphics[width=0.45\textwidth]{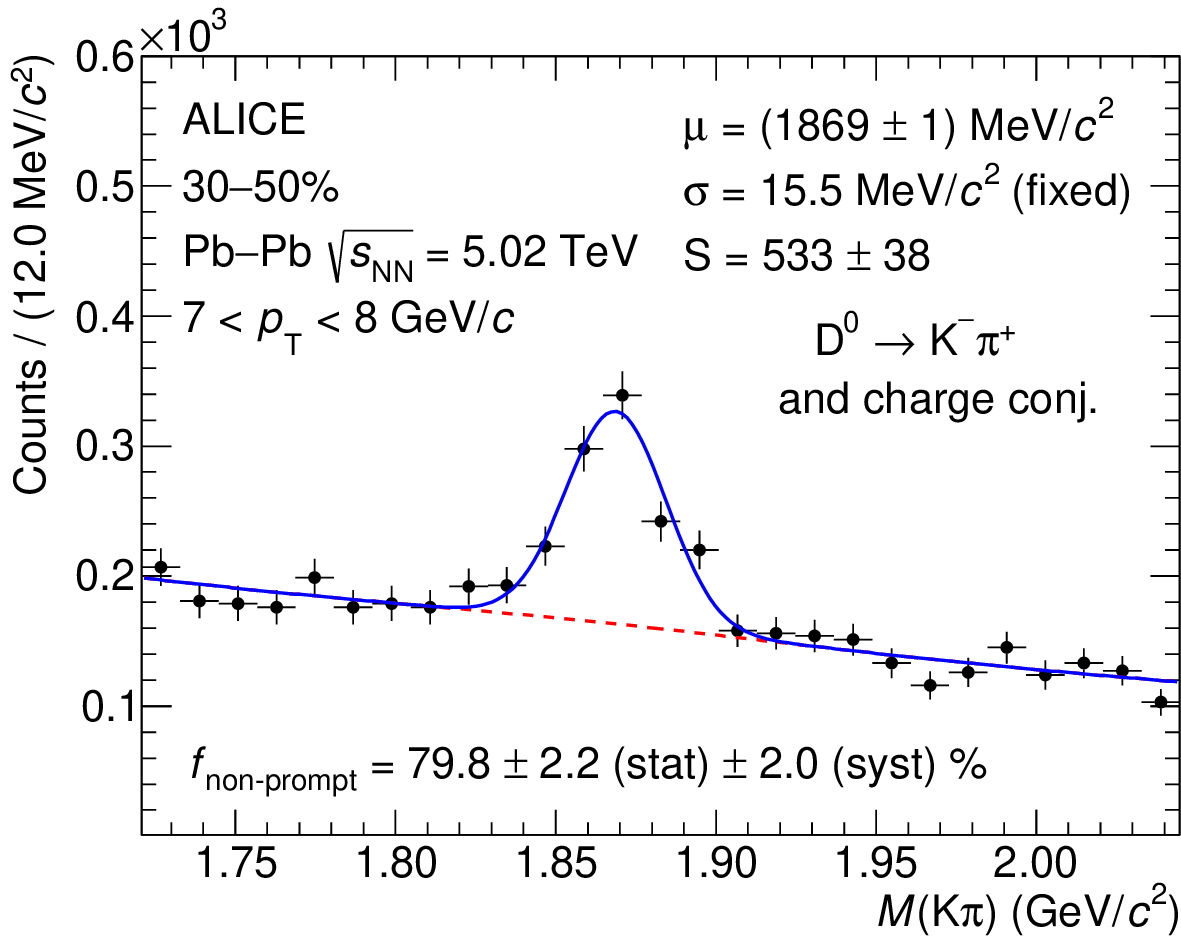}
\caption{Invariant-mass distributions for $\rm D^0$ candidates in selected $\pt$ intervals for the centrality class 0--10\% and 30--50\%. Fitted values for the $\rm D^0$ meson mass $\mu$, width $\sigma$, and raw yield $S$ are also given, the fraction of $\rm D^0$ candidates in the measured raw yield is reported with its statistical and systematic uncertainties.
Top row: non-prompt $\rm D^0$ mesons with $1<\pt<2~\gev/c$ in the 0--10\% centrality class, before (left) and after (right) subtraction of the background fit function.
Bottom row: non-prompt $\rm D^0$ mesons with $1<\pt<2~\gev/c$ (left) and $7<\pt<8~\gev/c$ (right) in the 30--50\% centrality class. 
}
\label{fig:D0mass}
\end{center}
\end{figure}

The raw yield was extracted in each $\pT$ interval via a binned maximum-likelihood fit to the candidate
invariant-mass distribution. The invariant-mass ($M$) distributions from which the raw yields with enhanced contribution of non-prompt $\rm D^0$ mesons are extracted are reported in Fig.~\ref{fig:D0mass}. The non-prompt $\rm D^0$ enriched invariant-mass distributions were fitted with a function composed of a Gaussian term for the signal and an exponential function to describe the background shape. For transverse-momentum $(\pt)$ intervals 2--6~$\gev/c$, a second-order polynomial function was used and below 2~$\gev/c$ a third-order polynomial function was used to describe the background shape. The contribution of signal candidates that are present in the invariant-mass distribution with the wrong decay-particle mass assignment (reflection) was parameterised by fitting the simulated reflection invariant-mass distributions with a double Gaussian function, and it was included in the total fit function. The ratio between the reflections and the signal yields was taken from simulations.  To improve the stability of the fits, the widths of the $\rm D^0$-meson signal peaks were fixed to the values extracted from data samples dominated by prompt candidates, given the naturally larger abundance of prompt compared to non-prompt $\rm D^0$ mesons.

The fraction of non-prompt ${\rm D}^0$ mesons in the raw yield $\fnonprompt$ was estimated by sampling the raw yield at different values of the BDT output related to the candidate probability of being a non-prompt ${\rm D}^0$ meson. In this way, a set of raw yields $\rawY{_\mathrm{i}}$ with different contributions of prompt and non-prompt ${\rm D}^0$ was obtained. The $\rawY{_\mathrm{i}}$ can be related to the corrected yields of prompt ($N_\mathrm{prompt}$) and non-prompt ($\Nnp$) ${\rm D}^0$ mesons via the acceptance-times-efficiency $(\rm Acc \times \epsilon)$ factors as follows
\begin{equation}
   \effP{\rm i}\times N_\mathrm{prompt} +  \effNP{\rm i}\times \Nnp - \rawY{\rm i} = \delta_\mathrm{i} .
\label{eq:eq_set}
\end{equation}
In the above equation, $\delta_\mathrm{i}$ represents a residual that accounts for the equation not holding exactly because of the uncertainties on $\rawY{\rm i}$, $\effNP{\rm i}$, and $\effP{\rm i}$. With $n\geq 2$ sets, starting from Eq.~\ref{eq:eq_set} a $\chi^2$ function can be defined, which can be minimised to obtain $N_\mathrm{prompt}$ and $\Nnp$. More details can be found in Ref.~\cite{Acharya:2021cqv}. However, rather than using the $\Nnp$ parameter from the $\chi^{2}$ minimisation, one of the $n$ sets with a high non-prompt component was selected as a working point (wp), and $\Nnp$ and $N_\mathrm{prompt}$ were used to calculate the $f_\mathrm{non\text{-}prompt\text{,wp}}$ fraction of the related raw yield $Y_{\rm{wp}}$. This choice facilitates the estimate of systematic uncertainties.
Then, to obtain the corrected non-prompt $\Dzero$-meson yield, the product $Y_{\rm{wp}}$$\times$$f_\mathrm{non\text{-}prompt\text{,wp}}$ was corrected for the corresponding acceptance-times-efficiency $\effNP{\rm wp}$ and divided by a factor $2\times{\rm{BR}}\times\Delta\pt\times\Delta{y}\times N_{\text{ev}}$, where $\Delta\pt$ and $\Delta{y}$ are the widths of the $\pt$ and rapidity intervals, BR is the branching ratio of the decay channel, $N_{\text{ev}}$ represents the number of analyzed events, and the factor 2 accounts for the fact that both particles and anti-particles are counted in the raw yield. 

\section{Systematic uncertainties}%
\label{sec:sys}

Several sources of systematic uncertainties on the non-prompt $\Dzero$-meson corrected yields were studied. The systematic uncertainty on the raw yield extraction ranges from 4\% to 14\% depending on $\pt$ and collision centrality class. It was evaluated by varying 
the lower and upper limits of the fit range, and the background fit function.
The contribution due to the uncertainty on track reconstruction efficiency (4--11\%) was evaluated 
by modifying the track-quality selections and by comparing the 
probability to prolong the TPC tracks to the ITS hits in data and simulation.
To estimate the uncertainty on the PID selection efficiency, the n$\sigma$ distributions for ``pure" samples of pions, kaons, and protons were compared in data and MC, and found to be compatible~\cite{ALICE:2021rxa}. Besides the analysis was repeated without PID selection, the resulting corrected yields were compatible with those obtained
with the PID selection and no systematic uncertainty was assigned.
The uncertainty on the selection efficiency (5--8\%) originates from imperfections in description in the MC simulation of the topological variables used for preselections and in the BDT. It was estimated by comparing the corrected yields obtained by repeating the analysis with different $\Dzero$-meson topological preselections, as well as the selection on the BDT response, resulting in a significant modification of the efficiency. The systematic uncertainty on the $f_\mathrm{non\text{-}prompt}$ fraction estimation (2--8\%) was evaluated by changing the selections defining the sets used for the $\chi^2$ minimisation calculation.
The systematic effect on the efficiency due to a possible difference between 
the real and simulated $\Dzero$-meson $\pt$ distributions (1--8\%)
was evaluated by reweighting the $\Dzero$-meson spectrum in the simulation so to match alternative $\pt$ distributions. 
The contributions of the different sources were summed in quadrature to obtain the total systematic uncertainty (8--19\%).

\section{Results}%
\begin{figure}[!t]
\centering
\includegraphics[width=.9\textwidth]{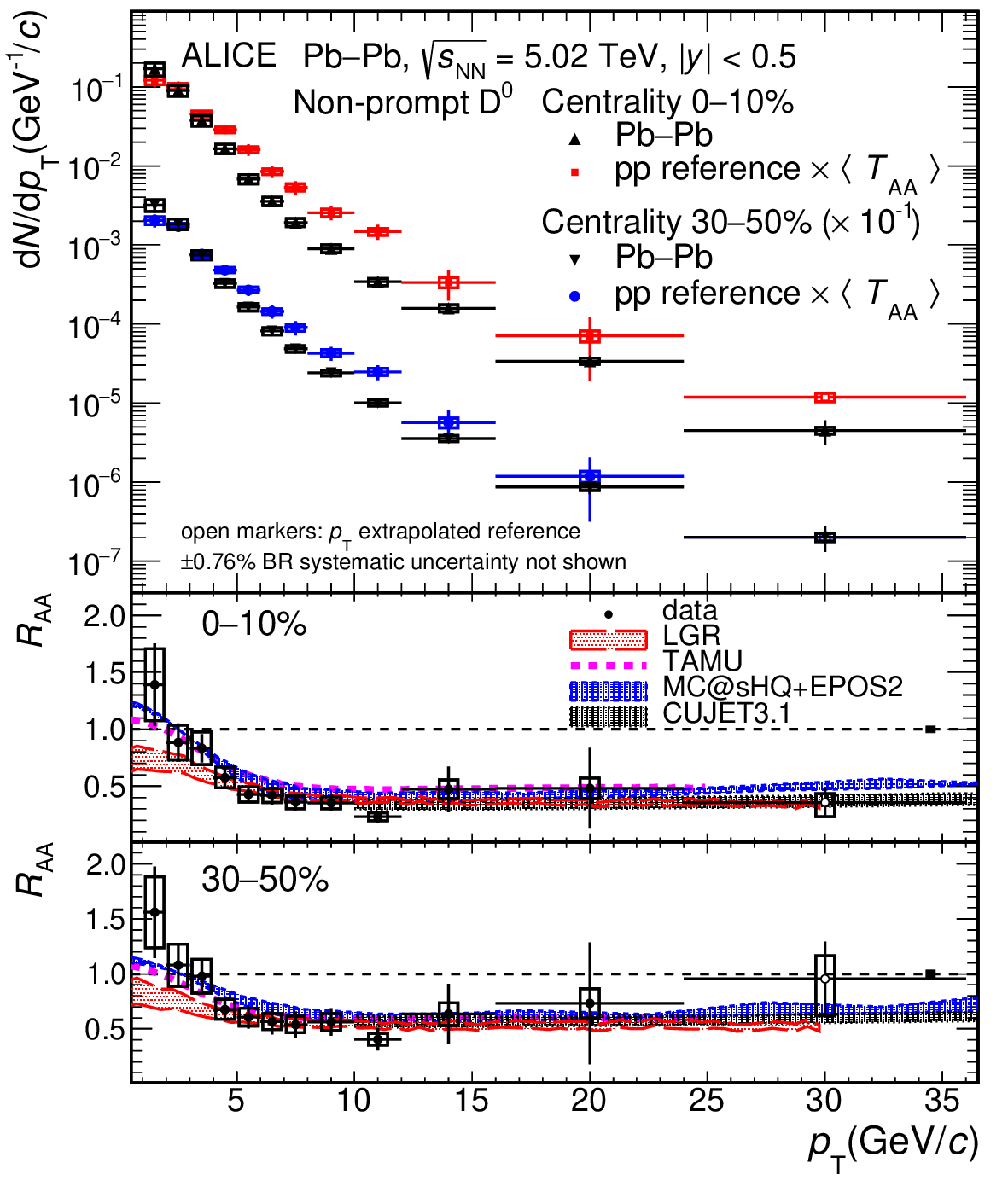}
\caption{Top panel: non-prompt $\rm D^0$-meson $\pt$-differential production yields in \PbPb{} collisions at \fivenn{} in the 0--10\% and 30--50\% centrality classes. 
The pp reference spectra, $\av{T_{\rm AA}}\times{\rm d}\sigma_{\rm pp}/{\rm d}\pt$~\cite{ALICE:2018tvk,Acharya:2021cqv}, are also shown.
Middle and bottom panels: $\pT$-differential $\RAA$ in the \cent{0}{10} (middle) and \cent{30}{50} (bottom) centrality classes, compared with model predictions~\cite{Nahrgang:2013xaa,Li:2020umn,Li:2019lex,He:2014cla,Shi:2019nyp}. 
Open markers indicate the points for which the pp reference is extrapolated (see text). Vertical bars, empty boxes, and the shaded box around $\RAA=1$ represent the statistical, systematic, and normalisation uncertainty, respectively. } 
\label{fig:RAA}
\end{figure}

\label{sec:resul}
The $\pT$-differential production yields of \np{} $\Dzero$ mesons in the 0--10\% and 30--50\% centrality classes in \PbPb{} collisions at \fivenn{}  for $\pt>1~\gev/c$ are shown in the top panel of Fig.~\ref{fig:RAA}.
They are compared to the corresponding pp reference cross section~\cite{Acharya:2021cqv} multiplied by $\av{T_{\rm AA}}$ in the given centrality range. For $24<\pT<36~\gev/c$, the pp reference cross section was extrapolated exploiting FONLL predictions in a similar way to that used in Refs.~\cite{Adam:2015sza,Acharya:2018hre}. To get an indication of the typical B-meson $\pt$ probed in the non-prompt $\Dzero$ $\pt$ intervals, a simulation was done in which ${\rm B}^{0}$ and ${\rm B}^{+}$ mesons were generated according to the $\pt$-differential spectrum expected from FONLL~\cite{Cacciari:1998it,Cacciari:2001td} and decayed with the PYTHIA 8.243 event generator. As an example, for non-prompt $\Dzero$ with $1<\pt<2$ ($10<\pt<12$)$~\gev/c$ the parent B-meson $\pt$ distribution has a median of $\pt\approx 3.3$ ($18.2$)$~\gev/c$ and an RMS of about $1.9$ ($6.2$)$~\gev/c$. Thus, the measured spectra probe B mesons down to $\pt$ lower than the B meson mass.


The $\RAA$ of \np{} $\Dzero$ mesons as a function of $\pT$ is shown in the middle and bottom panels of Fig.~\ref{fig:RAA} for the \cent{0}{10} and \cent{30}{50} centrality classes, respectively. The uncertainty on the $\RAA$ normalisation results from the quadratic sum of 
the pp normalisation uncertainty, 
the uncertainty on $\langle T_{\rm AA} \rangle$, and
the centrality interval definition uncertainty~\cite{Adam:2015sza}. The BR uncertainty is cancelled in the ratio, while all other sources are propagated as uncorrelated. 
For $\pt$ larger than about 5$~\gev/c$, the $\Raa$ does not change significantly with $\pT$ and it shows a suppression of the yields by a factor about $3$ ($2$) in the 
\cent{0}{10} (\cent{30}{50}) centrality class with respect to the pp reference scaled by $\av{T_{\rm AA}}$.
At lower $\pt$, the $\RAA$ increases with decreasing $\pT$. Within a $1\sigma$ uncertainty, it is compatible with unity in the interval $1<\pT<3~\gev/c$ ($1<\pT<4~\gev/c$) in the \cent{0}{10} (\cent{30}{50}) centrality class. Values above unity are slightly favoured by data in the range $1<\pT<2~\gev/c$. The measured $\RAA$ is compared with predictions from various models, namely \MCsHQ~\cite{Nahrgang:2013xaa}, LGR~\cite{Li:2020umn,Li:2019lex}, TAMU~\cite{He:2014cla}, and \CUJET~\cite{Shi:2019nyp}.
In the TAMU model, the heavy-quark interactions with the medium are described by elastic collisions only. The LGR, \MCsHQ{}, and \CUJET models include both radiative and collisional processes. The contribution of hadronisation via quark recombination, in addition to independent fragmentation, is considered in the TAMU, \MCsHQ{}, and LGR models.
All predictions describe the data within uncertainties in both centrality classes, except for TAMU, which tends to underestimate the suppression in the interval $5<\pt<12$~\GeVc in central collisions.
This comparison suggests that both radiative and collisional processes are important for beauty quark in-medium energy loss at LHC energies.

Shadowing and a modification of hadronisation can also modify the $\pt$-integrated yield of the final-state beauty hadrons, which is not influenced by the quark energy loss, and cause $\pt$-integrated $\RAA$ ($p_{\rm T} > 0$) to deviate from unity.
In order to test this, 
an extrapolation of the measured spectrum to the intervals $0<\pt<1~\gev/c$ was performed. The total non-prompt $\Dzero$-meson yields in $\abs{y}<0.5$ for $\pt>0$ in the 0--10\% and 30--50\% centrality classes are calculated by adding to the ``visible yields", computed by integrating in $\pt$ the $\pt$-differential yields measured for $\pt>1~\gev/c$, an estimate of the yield in $0<\pt<1~\gev/c$ reckoned as 
\begin{equation}
    \left.\frac{{\rm d}N}{{\rm d}\pt}\right |^{\rm non\text{-}prompt}_{\rm \text{\scriptsize Pb--Pb,\  extrap.}}(0<\pt<1~\gev/c)=R^{\rm prompt}_{\rm  AA,\ measured}\times\left. \frac{\Raa^{\rm non\text{-}prompt}}{\Raa^{\rm prompt}}\right|_{\rm model}\times\langle T_{\rm AA}\rangle\times\left.\frac{{\rm d}\sigma}{{\rm d}\pt}\right |^{\rm non\text{-}prompt}_{\rm  pp,\ extrap.}.
    \label{eq_extraYield01}
\end{equation}
In the above equation, all terms on the right side are evaluated in $0<\pt<1~\gev/c$. The value of the non-prompt $\Dzero$ cross section in pp collisions at $\sqrt{s}=5.02~\tev$, ${{\rm d}\sigma}/{{\rm d}\pt}$, is retrieved from Ref.~\cite{Acharya:2021cqv} (Tables~3 and 4) by scaling the cross section measured in $1<\pt<24~\gev/c$ by $1-\alpha=0.28^{+0.01}_{-0.04}$, where $\alpha$ represents the ratio of the cross section for $1<\pt<24~\gev/c$ to $\pt>0$ calculated using FONLL predictions~\cite{Cacciari:1998it,Cacciari:2001td}.
The contribution of non-prompt $\Dzero$ with $\pT>24~\gev/c$ to the total yield is below 0.1\% and significantly smaller than the uncertainty on the estimate of the yield in $0<\pt<1~\gev/c$, described later. Therefore, a correction to avoid the double counting of the contribution of the yield in the interval $24<\pt<36~\gev/c$, which is already accounted for in the visible Pb--Pb yield, as well as a specific extrapolation for $\pt>36~\gev/c$ were not considered necessary. In Eq.~\ref{eq_extraYield01} the pp cross section is multiplied by the nuclear overlap function $\langle T_{\rm AA}\rangle$ for the considered centrality interval and by an estimate of the non-prompt $\Dzero$-meson $\Raa$ obtained as the product of the measured prompt $\Dzero$-meson $\Raa$~\cite{ALICE:2021rxa} and an assumption for the ``double $\Raa$ ratio" $\Raa^{\rm non\text{-}prompt}/\Raa^{\rm prompt}$. For the latter, the $\pt$ shape of the prediction of the LGR model~\cite{Li:2020umn,Li:2019lex}, which describes the measured double $\Raa$ ratio for $\pt>1~\gev/c$ within uncertainties, is exploited. The model prediction is parametrised with a $5^{\rm th}$-order polynomial function, which is then used to fit the data in the interval $1<\pt<12~\gev/c$, leaving an overall scaling factor as the only free parameter of the fit. The value of the function at $\pt=0.5~\gev/c$ is assumed as the estimate of the double $\Raa$ ratio in $0<\pt<1~\gev/c$. The rescaling of the LGR prediction is performed mainly to avoid a potential unphysical discontinuity in the double $\RAA$ ratio between the measured and extrapolated ranges. It was verified that the original value of LGR at $\pt=0.5~\gev/c$ gives a value of the $\pt$-integrated yield that is compatible with that obtained with the default procedure within $1\sigma$ of the extrapolation uncertainty. The latter is obtained by summing in quadrature i) the statistical and systematic uncertainties on $R^{\rm prompt}_{\rm AA,\ measured}$, ii) the statistical and systematic uncertainties on ${{\rm d}\sigma}/{{\rm d}\pt}$, which include the uncertainty on the extrapolation factor $\alpha$ as well as the uncertainties on the visible cross section, and iii) the uncertainty on the double $\RAA$ ratio. The latter is determined by the sum in quadrature of the statistical uncertainty on the scaling factor of the LGR-based parametrisation of the double $\RAA$ ratio and the modeling uncertainty, which is determined from the envelope of the values obtained by reparametrising the double $\RAA$ ratio using the lower and upper predictions of LGR, as well as the TAMU~\cite{He:2014cla} model, which also reproduces the data within uncertainties for $\pt>1~\gev/c$. Moreover, also the values evaluated at $\pt=0.63~\gev/c$ rather than $\pt=0.5~\gev/c$ are considered, with the former value representing the average $\pt$ of non-prompt $\Dzero$ mesons with $0<\pt<1~\gev/c$ according to a simulation performed by decaying with PYTHIA 8.243~\cite{Sjostrand:2014zea} B mesons generated according to the expected $\pt$ spectrum of FONLL. The envelope spreads around the value of the double $\RAA$ ratio obtained with the default LGR prediction covering a relative variation of about ${}^{+19\%}_{-23\%}$ $({}^{+62\%}_{-37\%})$ in the \cent{0}{10} (\cent{30}{50}) centrality class. 

The systematic uncertainties on the visible yield are determined by summing those of the $\pt$-differential yields assuming that all uncertainty sources provide uncertainties correlated with $\pt$, with the exception of the yield-extraction uncertainties, which are assumed as uncorrelated with $\pt$ and summed in quadrature. The statistical uncertainty is calculated by summing in quadrature those on the $\pt$-differential yields.

The uncertainties on the visible yield and on the estimate of the yield in $0<\pt<1~\gev/c$ obtained with the procedure described above, are considered as uncorrelated in the sum performed to calculate the yield in $\pt>0$. The partial correlation induced by constraining the parametrisation of the double $\RAA$ ratio to the data is assumed to be negligible. Thanks to the low-$\pt$ reach of the measurement, the visible yields represent about 77\% and 82\% of the estimated total yields in the \cent{0}{10} and \cent{30}{50} centrality classes, respectively, and the extrapolation uncertainties are 10\% and 13\%, respectively.

The non-prompt $\Dzero$ yields estimated in the \cent{0}{10} and \cent{30}{50} centrality classes with the above procedure are divided by the non-prompt $\Dzero$ pp cross section for $\pt>0$~\cite{Acharya:2021cqv} scaled by the $\langle T_{\rm AA}\rangle$ value specific to each centrality class to get the non-prompt $\RAA$ for $\pt>0$. The uncertainty on the $\RAA$ is calculated taking properly into account that the $\alpha$ factor and the visible pp cross section are used for determining both the Pb--Pb and pp $\pt$-integrated yields. All other sources of uncertainties are considered uncorrelated between the Pb--Pb and pp yields, with exception of that on the BR, which cancels in the ratio.

The resulting total yields per unity of rapidity are $0.428\pm 0.033 \ ({\rm stat.})\pm 0.050\ ({\rm syst.}){ }^{+0.037}_{-0.042}\ ({\rm extr.})\pm 0.004\ ({\rm BR})$ and $0.079\pm 0.007 \ ({\rm stat.})\pm 0.009 \ ({\rm syst.}){ }^{+0.010}_{-0.007}\ ({\rm extr.})\pm 0.001 \ ({\rm BR})$ in the \cent{0}{10} and \cent{30}{50} centrality classes, respectively. The $\RAA$ for $\pt>0$ is $1.00\pm 0.10 \ ({\rm stat.})\pm 0.13 \ ({\rm syst.}){ }^{+0.08}_{-0.09} \ ({\rm extr.})\pm 0.02 \ ({\rm norm.})$ in the 0--10\% and $1.10\pm 0.12 \ ({\rm stat.})\pm 0.15 \ ({\rm syst.}){ }^{+0.14}_{-0.09} \ ({\rm extr.})\pm 0.03 \ ({\rm norm.})$ in the 30--50\% centrality class. Considering the statistical and systematic uncertainties, in both centrality classes, the $\RAA$ is compatible with unity within less than $1 \sigma$ and with the prompt $\Dzero$-meson $p_{\rm T}$-integrated $\RAA$~\cite{ALICE:2021rxa} within less than $1.5 \sigma$. 

\begin{figure}[h!]
\begin{center}
\includegraphics[width=0.9\textwidth]{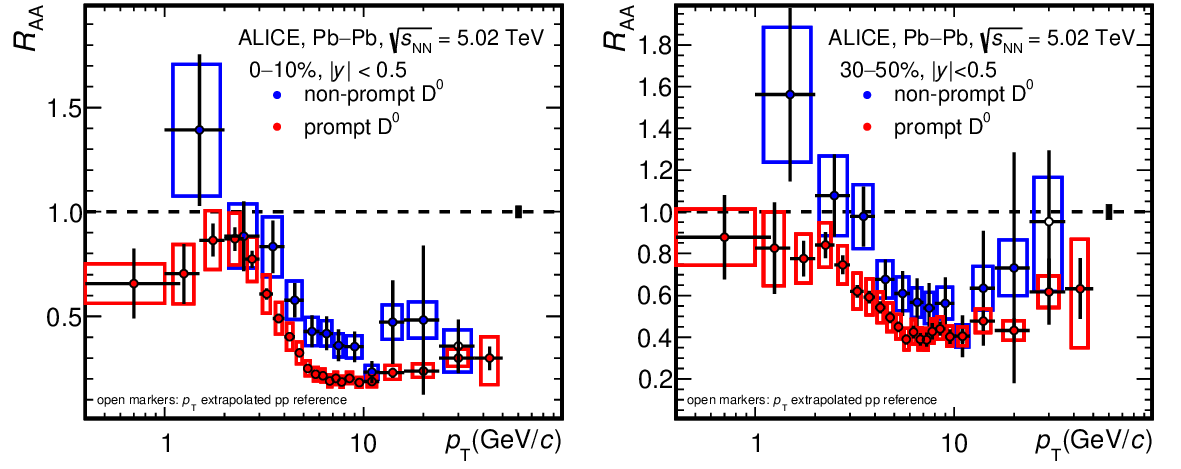}
\caption{Nuclear modification factor ($\RAA$) of non-prompt $\Dzero$ mesons in the centrality classes 0--10\% (left) and 30--50\% (right), compared with the $\RAA$ of prompt $\Dzero$ mesons~\cite{ALICE:2021rxa}. The statistical and total systematic uncertainties are shown as error bars and boxes, respectively. The normalisation uncertainties are shown as boxes around unity.}
\label{fig:p_np_raa}
\end{center}
\end{figure}

Figure~\ref{fig:p_np_raa} shows the $\RAA$ of non-prompt $\Dzero$ mesons in \PbPb{} collisions at \fivenn{}, compared with the $\RAA$ of prompt $\Dzero$ mesons~\cite{ALICE:2021rxa} in the 0--10\% and 30--50\% centrality classes. The non-prompt $\Dzero$ $\RAA$ is systematically higher than the prompt $\Dzero$ one for $\pt>5~\gev/c$ in both 0--10\% and 30--50\% centrality classes, indicating that non-prompt $\Dzero$ mesons are less suppressed than prompt $\Dzero$ ones and supporting the expectation that beauty quarks lose less energy than charm quarks because of their larger mass. 

\begin{figure}[h!]
\centering
\includegraphics[width=.9\textwidth]{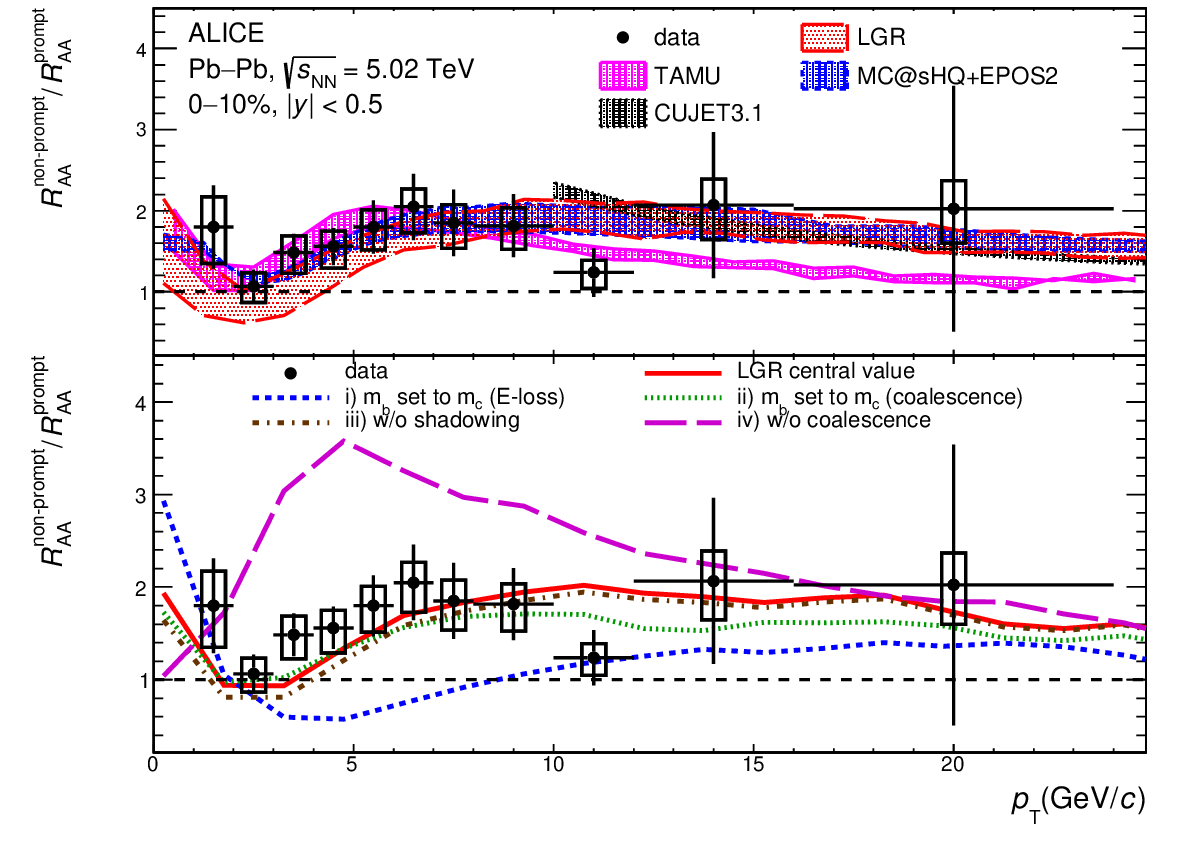} 
\caption{
Non-prompt to prompt~\cite{ALICE:2021rxa} $\Dzero$-meson $\RAA$ ratio
as a function of $\pt$ in the 0--10\% central \PbPb{} collisions at \fivenn{}, compared to  model predictions~\cite{Nahrgang:2013xaa,Li:2020umn,Li:2019lex,He:2014cla,Shi:2019nyp} (top), and to different modifications of LGR calculations (bottom).}
\label{fig:DRatio}
\end{figure}

The $R_{\rm AA}^{\text{\rm non-prompt}}/R_{\rm AA}^{\rm prompt}$ ratio 
as a function of $\pt$ is presented in Figure~\ref{fig:DRatio} for the 0--10\% central \PbPb{} collisions. In the computation of the ratio, the tracking-efficiency and normalisation uncertainties get cancelled. All other sources of systematic uncertainties were propagated as uncorrelated. 
As visible in the top panel, the $\pt$ trends of the $R_{\rm AA}^{\text{\rm non-prompt}}/R_{\rm AA}^{\rm prompt}$ ratio predicted by the LGR, \MCsHQ, and TAMU models at low $\pt$ are similar. They have a minimum close to unity in $2<\pt<3~\gev/c$ and increase towards lower and higher $\pt$, a trend resembling the data one, which however cannot be assessed in a conclusive way given the uncertainties. For $\pt>5~\gev/c$ the measured values do not vary significantly with $\pt$: their average is $1.70\pm0.18$, thus about $3.9\sigma$ above unity. All considered models, including CUJET3.1, predict a mild decrease of the ratio for $\pt\gsim10~\gevc$, which is steeper for CUJET3.1 and TAMU, with the latter predicting a maximum at $\pt\sim5~\gevc$. All models describe the data within uncertainties.

In the bottom panel of Fig.~\ref{fig:DRatio}, the ratio of the non-prompt to prompt $\rm D^0$-meson $\Raa$ is compared with predictions from the default LGR calculations as well as four different modifications of the LGR model:
i) using the charm-quark mass in the calculation of the beauty-quark energy loss,
ii) using the charm-quark mass in beauty-quark coalescence,
iii) excluding shadowing effects for both charm and beauty quarks, and
iv) excluding quark coalescence in both charm and beauty-quark hadronisation.
The configurations (ii) and (iii) give results similar to the default LGR calculation and can describe the data well.
The effect of shadowing is relevant mainly at low $\pt$ and it largely gets cancelled in the $\Raa$ ratio~\cite{Li:2020umn}. 
The usage of the charm-quark mass in beauty coalescence reduces the $\Raa$ ratio at high $\pt$, as expected from the reduced coalescence probability, while it has a marginal effect for $\pt\lsim7~\gev/c$. 
 By removing the quark recombination in hadronisation of both charm and beauty quarks (case iv), the $\RAA$ ratio is instead significantly enhanced for $\pt>1~\gev/c$ and reduced at lower $\pt$. This suggests that the minimum of the $\RAA$ ratio at $\pT\sim2.5~\gev/c$ in the default LGR calculations is mainly due to the formation of prompt D mesons via charm-quark coalescence. In this process, D mesons acquire a momentum larger than that of the parent charm quarks, causing a hardening of the prompt $\Dzero$-meson $\pt$ spectrum. 
By replacing the beauty-quark mass with that of the charm quark in the beauty-quark energy loss (case i), the $\RAA$ ratio reduces significantly for $\pT > 2.5~\gev/c$ and becomes lower than unity in $2<\pt<8~\gevc$, which is inconsistent with data. This supports the interpretation that the mass-dependence of quark in-medium energy-loss causes the $\RAA$ ratio to be significantly larger than unity at intermediate $\pT$. 

\section{Conclusions}%
\label{sec:con}

In summary, the $\RAA$ of \np{} $\Dzero$ mesons from beauty-hadron decays was measured at midrapidity, $\abs{y} < 0.5$, for $1<\pt<36~\gev/c$ in \PbPb collisions at \fivenn in the \cent{0}{10} and 30--50\% centrality classes. While $p_{\rm T}$-integrated $\RAA$ ($p_{\rm T} > 0$), which is not directly sensitive to partonic energy loss, is compatible with unity,
a significant suppression up to a factor of about three is observed for $\pT > 5~\gev/c$ in the 0--10\% central Pb--Pb collisions.
The data are described by models that include both collisional and radiative processes in the calculation of beauty quark in-medium energy loss and quark recombination as a hadronisation mechanism. 
The \np $\Dzero$-meson $\RAA$ is significantly larger than the prompt one. Models that describe their ratio as a function of $\pt$ encode a quark-mass dependence of energy loss, both at high $\pt$, where beauty quarks lose less energy than charm quarks via radiative processes, and at low $\pt$, a region in which collisional processes are more relevant and the interaction of heavy quarks with the medium can be described as a diffusion process.

\newenvironment{acknowledgement}{\relax}{\relax}
\begin{acknowledgement}
\section*{Acknowledgements}

The ALICE Collaboration would like to thank all its engineers and technicians for their invaluable contributions to the construction of the experiment and the CERN accelerator teams for the outstanding performance of the LHC complex.
The ALICE Collaboration gratefully acknowledges the resources and support provided by all Grid centres and the Worldwide LHC Computing Grid (WLCG) collaboration.
The ALICE Collaboration acknowledges the following funding agencies for their support in building and running the ALICE detector:
A. I. Alikhanyan National Science Laboratory (Yerevan Physics Institute) Foundation (ANSL), State Committee of Science and World Federation of Scientists (WFS), Armenia;
Austrian Academy of Sciences, Austrian Science Fund (FWF): [M 2467-N36] and Nationalstiftung f\"{u}r Forschung, Technologie und Entwicklung, Austria;
Ministry of Communications and High Technologies, National Nuclear Research Center, Azerbaijan;
Conselho Nacional de Desenvolvimento Cient\'{\i}fico e Tecnol\'{o}gico (CNPq), Financiadora de Estudos e Projetos (Finep), Funda\c{c}\~{a}o de Amparo \`{a} Pesquisa do Estado de S\~{a}o Paulo (FAPESP) and Universidade Federal do Rio Grande do Sul (UFRGS), Brazil;
Ministry of Education of China (MOEC) , Ministry of Science \& Technology of China (MSTC) and National Natural Science Foundation of China (NSFC), China;
Ministry of Science and Education and Croatian Science Foundation, Croatia;
Centro de Aplicaciones Tecnol\'{o}gicas y Desarrollo Nuclear (CEADEN), Cubaenerg\'{\i}a, Cuba;
Ministry of Education, Youth and Sports of the Czech Republic, Czech Republic;
The Danish Council for Independent Research | Natural Sciences, the VILLUM FONDEN and Danish National Research Foundation (DNRF), Denmark;
Helsinki Institute of Physics (HIP), Finland;
Commissariat \`{a} l'Energie Atomique (CEA) and Institut National de Physique Nucl\'{e}aire et de Physique des Particules (IN2P3) and Centre National de la Recherche Scientifique (CNRS), France;
Bundesministerium f\"{u}r Bildung und Forschung (BMBF) and GSI Helmholtzzentrum f\"{u}r Schwerionenforschung GmbH, Germany;
General Secretariat for Research and Technology, Ministry of Education, Research and Religions, Greece;
National Research, Development and Innovation Office, Hungary;
Department of Atomic Energy Government of India (DAE), Department of Science and Technology, Government of India (DST), University Grants Commission, Government of India (UGC) and Council of Scientific and Industrial Research (CSIR), India;
Indonesian Institute of Science, Indonesia;
Istituto Nazionale di Fisica Nucleare (INFN), Italy;
Japanese Ministry of Education, Culture, Sports, Science and Technology (MEXT) and Japan Society for the Promotion of Science (JSPS) KAKENHI, Japan;
Consejo Nacional de Ciencia (CONACYT) y Tecnolog\'{i}a, through Fondo de Cooperaci\'{o}n Internacional en Ciencia y Tecnolog\'{i}a (FONCICYT) and Direcci\'{o}n General de Asuntos del Personal Academico (DGAPA), Mexico;
Nederlandse Organisatie voor Wetenschappelijk Onderzoek (NWO), Netherlands;
The Research Council of Norway, Norway;
Commission on Science and Technology for Sustainable Development in the South (COMSATS), Pakistan;
Pontificia Universidad Cat\'{o}lica del Per\'{u}, Peru;
Ministry of Education and Science, National Science Centre and WUT ID-UB, Poland;
Korea Institute of Science and Technology Information and National Research Foundation of Korea (NRF), Republic of Korea;
Ministry of Education and Scientific Research, Institute of Atomic Physics, Ministry of Research and Innovation and Institute of Atomic Physics and University Politehnica of Bucharest, Romania;
Joint Institute for Nuclear Research (JINR), Ministry of Education and Science of the Russian Federation, National Research Centre Kurchatov Institute, Russian Science Foundation and Russian Foundation for Basic Research, Russia;
Ministry of Education, Science, Research and Sport of the Slovak Republic, Slovakia;
National Research Foundation of South Africa, South Africa;
Swedish Research Council (VR) and Knut \& Alice Wallenberg Foundation (KAW), Sweden;
European Organization for Nuclear Research, Switzerland;
Suranaree University of Technology (SUT), National Science and Technology Development Agency (NSDTA), Suranaree University of Technology (SUT), Thailand Science Research and Innovation (TSRI) and National Science, Research and Innovation Fund (NSRF), Thailand;
Turkish Energy, Nuclear and Mineral Research Agency (TENMAK), Turkey;
National Academy of  Sciences of Ukraine, Ukraine;
Science and Technology Facilities Council (STFC), United Kingdom;
National Science Foundation of the United States of America (NSF) and United States Department of Energy, Office of Nuclear Physics (DOE NP), United States of America.
\end{acknowledgement}

\bibliographystyle{utphys}
\bibliography{AlinpD0PbPb}

\newpage
\appendix

\section{The ALICE Collaboration}
\label{app:collab}
\small
\begin{flushleft} 


S.~Acharya$^{\rm 143}$, 
D.~Adamov\'{a}$^{\rm 97}$, 
A.~Adler$^{\rm 75}$, 
J.~Adolfsson$^{\rm 82}$, 
G.~Aglieri Rinella$^{\rm 35}$, 
M.~Agnello$^{\rm 31}$, 
N.~Agrawal$^{\rm 55}$, 
Z.~Ahammed$^{\rm 143}$, 
S.~Ahmad$^{\rm 17}$, 
S.U.~Ahn$^{\rm 77}$
% ...

\bigskip

\bigskip 

\textbf{\Large Affiliation Notes}

\bigskip 

% ...
$^{\rm VII}$ Also at: University of Kansas, Lawrence, Kansas, United States\\

\bigskip

\bigskip 

\textbf{\Large Collaboration Institutes}

\bigskip 

% ...
$^{17}$ Department of Physics, Aligarh Muslim University, Aligarh, India\\
% ...
$^{31}$ Dipartimento DISAT del Politecnico and Sezione INFN, Turin, Italy\\
% ...
$^{35}$ European Organization for Nuclear Research (CERN), Geneva, Switzerland\\
% ...
$^{55}$ INFN, Sezione di Bologna, Bologna, Italy\\
% ...
$^{75}$ Johann-Wolfgang-Goethe Universit\"{a}t Frankfurt Institut f\"{u}r Informatik, Fachbereich Informatik und Mathematik, Frankfurt, Germany\\
% ...
$^{77}$ Korea Institute of Science and Technology Information, Daejeon, Republic of Korea\\
% ...
$^{82}$ Lund University Department of Physics, Division of Particle Physics, Lund, Sweden\\
% ...
$^{97}$ Nuclear Physics Institute of the Czech Academy of Sciences, \v{R}e\v{z} u Prahy, Czech Republic\\
% ...
$^{143}$ Variable Energy Cyclotron Centre, Homi Bhabha National Institute, Kolkata, India\\
% ...


\end{flushleft} 

\end{document}